\documentclass[a4wide]{article}

\usepackage{epsfig}
\usepackage{amsfonts}
\usepackage{cite}

\newcommand{\vb}[1]{{\mathbf{#1}}}
\newcommand{\lb}[1]{\label{#1}}

\newcommand{\bc}{\begin{center}}
\newcommand{\ec}{\end{center}}
\newcommand{\be}{\begin{equation}}
\newcommand{\ee}{\end{equation}}
\newcommand{\bea}{\begin{eqnarray}}
\newcommand{\eea}{\end{eqnarray}}
\newcommand{\ba}[1]{\begin{array}{#1}}
\newcommand{\ea}{\end{array}}

\newcommand{\bt}[1]{\begin{table}[ht]\centering\begin{tabular}{#1}}
\newcommand{\et}[1]{\end{tabular}\caption{\small#1}\end{table}}
\newcommand{\fig}[3]{\begin{figure}[htb]\epsfxsize=80mm\bigskip\centerline{\epsfbox{#1}}\caption{\small\it #2 \label{#3}}\bigskip\end{figure}}

\begin{document}

\thispagestyle{empty}

\begin{center}

\vspace{1.5 truecm}

{\large\bf{QCD Corrections to QED Vacuum Polarization}}
\vspace{2 truecm}\\

{\bf P. Castelo Ferreira\\ \tt pedro.castelo.ferreira@ist.utl.pt}\\[5mm]

{\bf J. Dias de Deus\\ \tt jdd@fisica.ist.utl.pt}\\[5mm]

{\small{\it CENTRA, Instituto Superior T\'ecnico, Av. Rovisco Pais, 1049-001 Lisboa, Portugal}}\\[2cm]

{\bf\sc Abstract}\\[5mm]

\begin{minipage}{12cm}
We compute QCD corrections to QED calculations for vacuum polarization in background magnetic fields.
Formally, the diagram for virtual $e\bar{e}$ loops is identical to the one for virtual $q\bar{q}$ loops.
However due to confinement, or to the growth of $\alpha_s$ as $p^2$
decreases, a direct calculation of the diagram is not allowed. At large
$p^2$ we consider the virtual $q\bar{q}$ diagram, in the intermediate region
we discuss the role of the contribution of quark condensates
$\left\langle q\bar{q}\right\rangle$ and at the low-energy limit we consider the
$\pi^0$, as well as charged pion $\pi^+\pi^-$ loops. Although these effects
seem to be out of the measurement accuracy of photon-photon laboratory experiments they may be
relevant for $\gamma$-ray burst propagation. In particular, for emissions from the center of
the galaxy ($8.5$~kpc), we show that the mixing between the neutral pseudo-scalar pion $\pi_0$ and photons renders
a deviation from the power-law spectrum in the $TeV$ range. As for scalar quark condensates $\left\langle q\bar{q}\right\rangle$
and virtual $q\bar{q}$ loops are relevant only for very high radiation density $\sim 300\,MeV/fm^3$
and very strong magnetic fields of order $\sim 10^{14}\,T$.
\end{minipage}

\end{center}

\vfill
\begin{flushleft}
\end{flushleft}

\newpage
\section{Introduction}

In the presence of background electromagnetic fields second order QED corrections in the fine-structure constant $\alpha$
to vacuum polarization due to quantum vacuum oscillations, i.e. electron-positron virtual loops~\cite{HE,Schwinger},
is a well studied subject, including Delbr\"uck Scattering~\cite{Delbruck}, photon-splitting~\cite{Adler,Birula},
photon-photon interactions~\cite{lund,mark} and to semi-classical interactions with pseudo-scalar particles~\cite{axion1,axion2,axion3}.
QED effects are well established, D\"elbruck scattering has been experimentally observed in light scattering by heavy
nuclei~\cite{Jarlskog} as well as contributing in second order in $\alpha Z$ to the Lamb shift~\cite{Lamb}. However, a
direct signature for interaction with axion-like pseudo-scalars has been consistently
verified~\cite{PVLA,Roncadelli}. Independent experiments to detect this sort of interaction have now
been considered~\cite{Gies,Tito}. In addition in astro-particle observations, due to the wide range of energies accessible,
QED effects are the main contribution for the optical depth. Such effects include particle production
(also known as photon desintegration)~\cite{desintegration}, vacuum polarization and photon splitting~\cite{splitting}.
Also astro-particle observations seem to be the most promising candidate to test
pseudo-scalar interactions~\cite{gamma_conv,pseudo_ray,Raffelt} trough, for example, analysis of gamma-ray burst
conversion rates~\cite{HESS} and its polarization characteristics~\cite{gamma_pol}.

In this work we study the second order perturbative corrections in the fine-structure constant due to
the strong interactions. Namely we analyse quark loops, quark condensates and meson
contributions to vacuum polarization. The second order contribution to the polarization of the vacuum due
to electron-positron virtual pair production is given by the Euler-Heisenberg Lagrangian~\cite{HE,Schwinger}
\be
\ba{rcl}
{\mathcal{L}}^{(2)}_{e\bar{e}}&=&\xi_e\left[4\left(F_{\mu\nu}F^{\mu\nu}\right)^2+7\left(\epsilon^{\mu\nu\delta\rho}F_{\mu\nu}F_{\delta\rho}\right)^2\right]\ ,\\[5mm]
\xi_e&=&\displaystyle\frac{2\alpha}{45\,(B^e_c)^2}=1.32\times 10^{-24}\,T^{-2}\ ,\\[5mm]
B^e_c&=&\displaystyle\frac{m_e^2 c^2}{e\,\hbar}\ .
\ea
\lb{L_e}
\ee
The respective vacuum polarization dispersion relation for radiation
propagating in vacuum under an orthogonal magnetic field $B_0\ll B^e_c$ is
expressed in terms of the eigenvalues $\lambda^e_\perp$ and $\lambda^e_\parallel$, respectively orthogonal
and parallel to the magnetic field~\cite{Birula,Adler},
\be
\ba{rcl}
\omega_{\perp,\parallel}&=&k\left(1-\lambda^e_{\perp,\parallel}B_0^2\right)\ ,\\[5mm]
\lambda^e_\perp&=&8\xi_eB_0^2\ ,\ \lambda^e_\parallel=14\xi_eB_0^2\ .
\ea
\lb{Pi2_ee}
\ee
Also within the framework of QED we have the contribution of other fermionic loops than
the ones due to the electron. Relevant to the present study we have the muon loops $\mu\bar{\mu}$
which give the contribution
\be
\ba{rcl}
\lambda^\mu_{\perp,\parallel}&=&\Delta\xi_\mu\,\lambda^e_{\perp,\parallel}\ ,\\[5mm]
\Delta\xi_\mu&=&\displaystyle\frac{\xi_\mu}{\xi_e}=\left(\frac{m_e}{m_\mu}\right)^4=5.43\times 10^{-10}\ ,
\ea
\lb{Pi2_mumu}
\ee
being of the same order of magnitude of the QCD corrections that we are addressing here.
Due to its higher mass the $\tau$ gives a contribution five orders of magnitude lower
($\sim 6\times 10^{-15}$), hence not being relevant here.

\section{QCD Contributions}

Naively, we can expect that the same kind of physics applies to quark-antiquark virtual pair production. In the
presence of an external field we have in general the diagram of figure~\ref{fig.qq}.
\fig{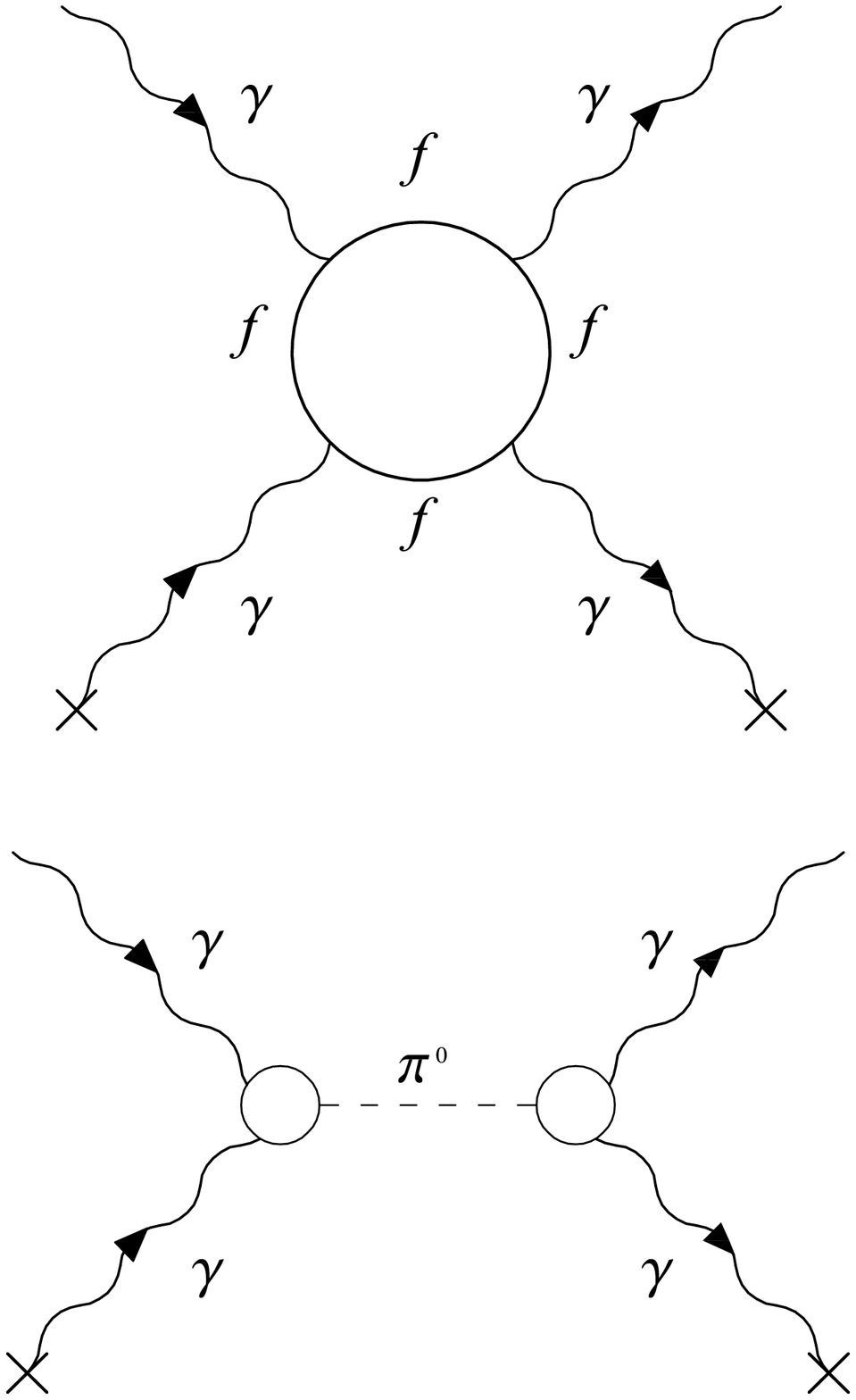}{The diagrams for fermion-antifermion loops and the
exchange of a $\pi^0$ neutral meson. The vertex $\pi^0\gamma\gamma$
includes the axial anomaly.}{fig.qq}
We write in the case of $q\bar{q}$ virtual pair production in order $\alpha^2$ the Euler-Heisenberg
Lagrangian
\be
\ba{rcl}
{\mathcal{L}}^{(2)}_{q\bar{q}}&=&\xi_q\left[4\left(F_{\mu\nu}F^{\mu\nu}\right)^2+7\left(\epsilon^{\mu\nu\delta\rho}F_{\mu\nu}F_{\delta\rho}\right)^2\right]\ ,\\[5mm]
\xi_q&=&\displaystyle\delta_q\frac{2\alpha Q_q^2}{45\,(B^q_c)^2}\ ,\ B^q_c=3\frac{m_q^2 c^2}{e\,Q_q\,\hbar}\ .
\ea
\lb{L_q}
\ee
The factor of $3$ comes from the summation over colours and $Q_q$ stands for the quark fractional charge.
As for the quark masses $m_q$ correspond to the \textit{renormalized} masses that appear in the quark propagator.
Under an external magnetic field $B_0$ the relation of the polarization due to $q\bar{q}$ with the polarization
due to $e\bar{e}$ corresponding to electron-positron loops is given in terms of the corrections to the parallel
and orthogonal vacuum dispersion relation eigenvalues $\lambda^q_\perp$ and $\lambda^q_\parallel$ by
\be
\ba{rcl}
\lambda^q_{\perp,\parallel}&=&\Delta\xi_q\,\lambda^e_{\perp,\parallel}\ ,\\[5mm]
\Delta\xi_q&=&\displaystyle\frac{\xi_q}{\xi_e}=3\,\delta_q\,\left(\frac{m_e\,Q_q}{m_q}\right)^4\approx 6.41\,\delta_q\times 10^{-4}\ .
\ea
\lb{Pi2_qq}
\ee
Here we consider the up quark mass $m_q\approx 5\,MeV$ and charge $Q_q=2/3$ and
$\delta_q=w_{\Lambda_q}/w_{\mathrm{tot}}<1$ is a phase space correction due
to confinement of strong interactions. We know that at low energies
there are no free quarks, therefore quark loops carrying small
momenta cannot be considered. The way out is to introduce a lower
cut-off $\Lambda_q$ in the loop momenta such that only the higher momenta
contribution to the loop is considered.

The probability for the full range of momenta (i.e. $p^2\in]0,+\infty[$) is given by the series~\cite{IZ}
\be
w_{\mathrm{tot}}=\frac{\alpha\,\vb{B}^2}{\pi^2}\sum_{n=1}^{\infty}\frac{1}{n^2}e^{-\frac{n\pi m_q^2}{|Q_q\vb{B}|}}\ .
\lb{w_tot}
\ee
Due to confinement and the increase of $\alpha_s$ for small values of $p^2$, we introduce a cut-off $\Lambda_q$ that
truncates the series~(\ref{w_tot}) by excluding the low $p^2$ region
\be
w_{\Lambda_q}=\frac{\alpha\,\vb{B}^2}{\pi^2}\sum_{n=n_{\Lambda_q}}^{\infty}\frac{1}{n^2}e^{-\frac{n\pi m_q^2}{|Q_q\vb{B}|}}\ ,
\lb{w_L}
\ee
obtaining the relation
\be
n_{\Lambda_q}=\left(\frac{\Lambda_q}{m_q}\right)^2\ .
\ee
For the light quarks (summing over up and down quark masses) with mass of order $m_q\sim 10\,MeV$~\cite{PDG}
we have $n_{\Lambda_q}\sim 3600$ such that $\delta_q\sim 10^{-10^{14}/|B|}$. This value is obtained from the
leading term contribution from the above series for $\omega_{\Lambda_q}$. Here we considered
$\Lambda_q\sim 600\,MeV$, this being the value for which the strong interactions
coupling constant becomes unity, $\alpha_s\sim 1$~\cite{QCD}, such that
below this energy threshold, QCD is in a non-perturbative regime.
The free quark loop contribution to vacuum polarization is therefore
negligible. This contribution will only be relevant
for very strong magnetic fields of order $B\sim 10^{14}\,T$.

The low-energy quark states (corresponding to the light mesons) are the $\pi$'s.
In low energy physics these particles can be used
as \textit{fundamental} bosons within the framework of
Chiral Perturbation Theory (ChPT)~\cite{ChPT}. Therefore
below the cut-off $p^2<\Lambda_q$ the main contribution
is due to the neutral meson $\pi^0$ with an effective scalar
Lagrangian~\cite{Schwinger}:
\be
\ba{rcl}
{\mathcal{L}}^{(2)}_{\pi^0}&=&\displaystyle \frac{1}{4}g_{\pi\gamma\gamma}\,\phi_{\pi^0}\,\epsilon^{\mu\nu\lambda\rho}F_{\mu\nu}F_{\lambda\rho}\ ,\\[5mm]
g_{\pi\gamma\gamma}&=&\displaystyle\frac{\alpha}{\pi f_\pi}=2.49\times 10^{-2}\,GeV^{-1}\ ,
\ea
\lb{L_pi0}
\ee
where the coupling $g_{\pi\gamma\gamma}$ is taken from the Adler-Bell-Jackiw anomaly coefficient~\cite{ABJ}
and the pion decay constant is taken to be $f_\pi=93\,MeV$.
The respective contribution to the dispersion relation of radiation traveling in vacuum
(corresponding to the $\pi$ diagram of figure~\ref{fig.qq}) is
\be
\ba{rcl}
\lambda^{\pi^0}_\perp&=&0\ ,\ \lambda^{\pi^0}_\parallel=\Delta\xi_{\pi^0}\lambda^e_\parallel\ ,\\[5mm]
\Delta\xi_{\pi^0}&=&\displaystyle\frac{g_{\pi^0\gamma\gamma}^2}{14\xi_e}=\frac{45\,m_e^4}{14\pi^2\,m_\pi^2\,f_\pi^2}=1.40\times 10^{-10}\ .
\ea
\lb{Pi2_pi0}
\ee
We considered the pion mass $m_\pi=135\,MeV$, for higher masses the contributions are
for most applications negligible when compared to the $\pi^0$ effect.

In addition we can have loops of the lighter charged mesons
$\pi^+\pi^-$ whose Euler-Heisenberg Lagrangian reads~\cite{Schwinger}
\be
\ba{rcl}
{\mathcal{L}}^{(2)}_{\pi^+\pi^-}&=&\xi_{\pi^\pm}\left[7\left(F_{\mu\nu}F^{\mu\nu}\right)^2+4\left(\epsilon^{\mu\nu\delta\rho}F_{\mu\nu}F_{\delta\rho}\right)^2\right]\ ,\\[5mm]
\xi_{\pi^\pm}&=&\displaystyle\frac{\alpha}{45\,(B^e_c)^2}\ ,\ B^e_c=\frac{m_\pi^2 c^2}{e\,\hbar}\ ,
\ea
\lb{L_pipm}
\ee
hence contributing a correction to the vacuum dispersion relation
of approximately the same order of magnitude given by
\be
\ba{rcl}
\lambda^{\pi^\pm}_{\perp}&=&\displaystyle\frac{7}{4}\Delta\xi_{\pi^\pm}\,\lambda^e_{\perp}\ ,\ \lambda^{\pi^\pm}_{\parallel}=\frac{4}{7}\Delta\xi_{\pi^\pm}\,\lambda^e_{\parallel}\\[5mm]
\Delta\xi_{\pi^\pm}&=&\displaystyle\frac{\xi_\pi}{\xi_e}=\frac{1}{2}\left(\frac{m_e f_\pi}{m_\pi^2}\right)^4=2.29\times 10^{-11}\ .
\ea
\lb{Pi2_pipm}
\ee

There is yet another contribution that we can consider. In the presence of background
magnetic fields there is a vacuum polarization contribution due to
quark condensates. Within the Schwinger-Euler-Heisenberg formalism~\cite{HE,Schwinger} in the context
of ChPT~\cite{QCD_vac_ChPT} a correction to the vacuum dispersion is obtained
\be
\ba{rcl}
\lambda^c_\perp&=&\Delta\xi_c\,\lambda^e_\perp\ ,\ \lambda^c_\parallel=0\ ,\\[5mm]
\Delta\xi_c&=&\displaystyle\frac{\xi_c}{8\xi_e}=\frac{15\,m_e^4}{128\,f_\pi^4}\ln\left(\frac{\Lambda^2}{m_\pi^2}\right)\\[5mm]
           &=&\displaystyle 1.05\times 10^{-10}\,\ln\left(\frac{\Lambda^2}{m_\pi^2}\right)\ .
\ea
\lb{Pi2_c}
\ee
Taking the quark condensate ultra-violet cut-off $\Lambda\approx 300\,MeV$ we obtain that
$\Delta\xi_c\approx 1.69\times 10^{-10}$. Next we give some details on how quark condensates
are obtained and explain which regimes exist depending on the loop momentum. The parallel
vacuum polarization for ChPT is given by the integral
\be
\ba{rcl}
\Pi_{\left\langle q\bar{q}\right\rangle}&=&\displaystyle\int_0^\infty ds\, I_{\left\langle q\bar{q}\right\rangle}\ ,\\[5mm]
I_{\left\langle q\bar{q}\right\rangle}&=&\displaystyle-\frac{\alpha\,B}{12\,f_\pi^4}\,\frac{1}{s^2}\left[\alpha\,B\,\cot(\alpha\,B\,s)-\frac{1}{s}\right]\ .
\ea
\lb{int_qq}
\ee
This distribution is represented in figure~\ref{fig.poles}.
\fig{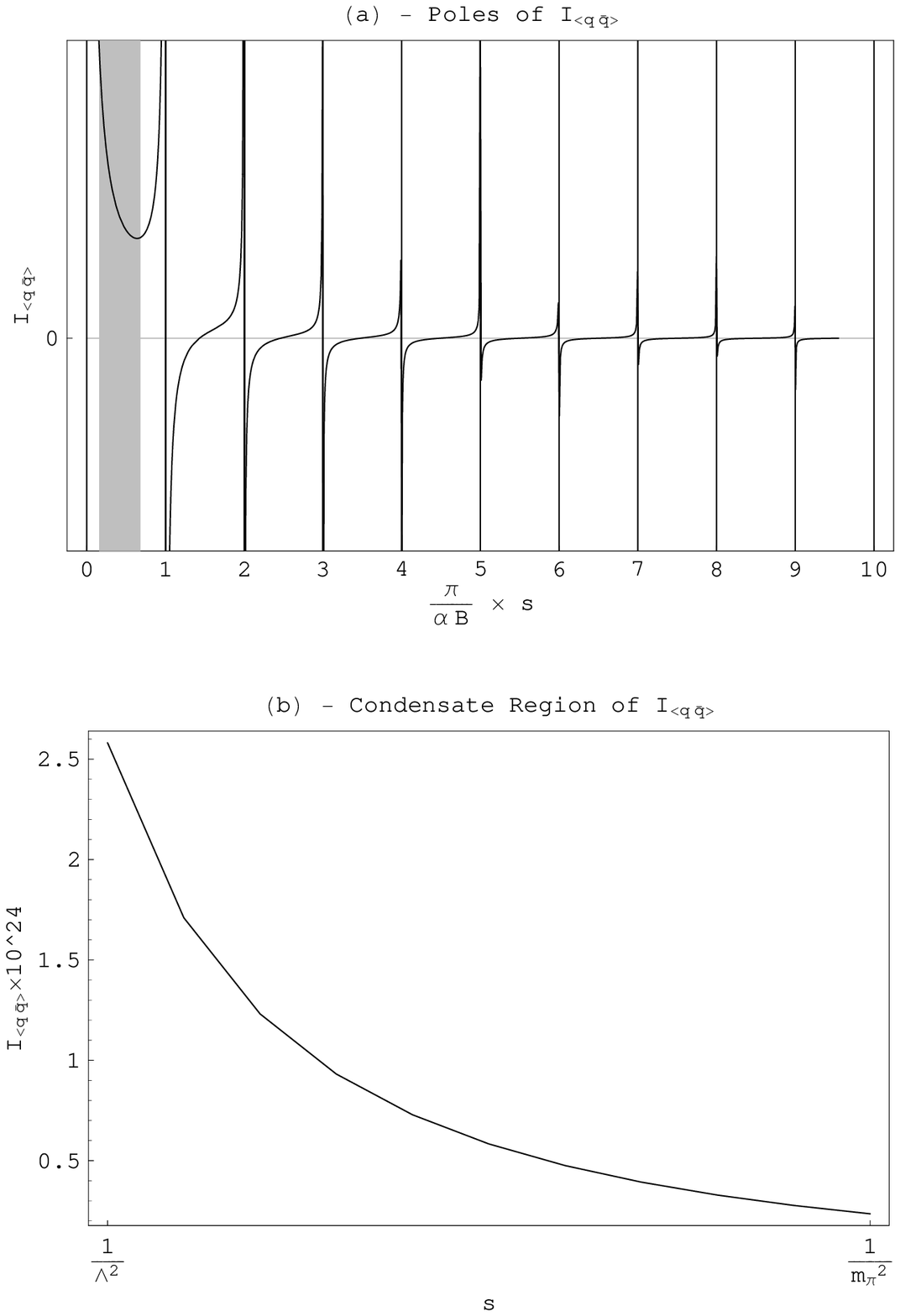}{(a) The integrand~(\ref{int_qq}). The poles at $s=(n-1) \pi/\alpha B$ (for $n=1,2,\ldots,\infty$)
are marked by vertical lines and contribute to the pion vacuum polarization. (b) The same integrand between both cut-offs
$m_\pi=135\,MeV$ and $\Lambda=300\,MeV$ for $B=5.5\,T$. It corresponds to the marked region between the poles at $s=0$ and
$s=\pi/\alpha B$ of (a).}{fig.poles}
The contributions considered here are due to the poles below the cut-off $s<1/\Lambda^2$.
For weak fields the only pole that contributes for pion loops is at $s=0$. It corresponds to the
$\pi^+\pi^-$ loops and the relative magnitude of its effect has already
been discussed and is given in equation~(\ref{Pi2_pipm}). Above the cut-off $s>1/\Lambda^2$ we consider the
quark loops instead of the meson distributions. The novel interesting feature in this framework is that we
have a new contribution between the pole $s=0$ and $s=\pi/\alpha|B|$, it corresponds to the
quark condensate. We note that from a more fundamental level based in Nambu-Jona-Lasinio theory~\cite{NLJ}
the quark condensates contribution is of the same order of magnitude~\cite{QCD_vac_NLJ}. There is an important point to stress,
NJL consider explicit actions for the quarks instead of the effective actions for
the mesons considered in ChPT, the condensate cut-off $\Lambda$ should correspond
in NJL to the confinement energy. These theories were originally motivated by
superconductivity and the relation between ChPT and NJL is equivalent
to the relation between Landau-Ginzburg effective theory~\cite{LG} and
the Bardeen-Cooper-Schriffer microscopic theory~\cite{BCS}
for superconductivity.

We summarize the allowed effects and their magnitude for several
ranges of loop momenta $p$ in table~\ref{table_lambda}.
\begin{table}
\bea
\ba{c|ccc}
p\ (MeV)&\Delta\xi_{\pi^0}&\Delta\xi_c&\Delta\xi_q\\ \hline
        &                 &        & \\[5mm]
>600    &\ 0              &\ 0        &\displaystyle\ 10^{(-10^{14})}\\[5mm]
140-600 &\ 0              &\displaystyle\ 9.67\times 10^{-11}&\ 0\\[5mm]
<140    &\displaystyle\ 1.40\times 10^{-10}&\ 0& 0
\ea
\nonumber
\eea
\caption{The several QCD effects in the presence of weak fields and their magnitude for
the several ranges of the loop momenta $p$. To exist, the quark condensate requires
a high density of energy.\lb{table_lambda}}
\end{table}

We note that the $\pi_0$ is a $0^{-+}$ pseudo-scalar such that $CP$-symmetry is conserved,
the Lagrangian~(\ref{L_pi0}) is a scalar. More generally we may consider also the
contributions of other $0^{-+}$ pseudo-scalars as the $\eta$'s, however their masses are
higher than the mass of the pion (the lighter being $m_\eta\approx 547\,MeV$),
hence their contribution is negligible by several orders of magnitude.

As for quark condensates are $0^{++}$ scalars. Although we have already present the effects of these condensates
in equation~(\ref{Pi2_c}) obtained within the mean-field framework of ChPt, we note that we can recast this
effects diagrammatically considering the scattering of photons by an intermediate scalar. However if one demands
$CP$-invariance the effective action is expressed as $\mathcal{L}=g_{\left\langle q\bar{q}\right\rangle}\phi_{\left\langle q\bar{q}\right\rangle}F^{mu\nu}F_{\mu\nu}$, where $\phi_{\left\langle q\bar{q}\right\rangle}$ represents now a scalar.
In the same manner one may consider the other $0^{++}$ scalar mesons $f_0$ and $a_0$~\cite{mesons}. Although these processes are
allowed and the couplings are of the same magnitude of the pion, are negligible due to much higher masses~\cite{PDG}.

In addition we also note that the quark condensates may only exist when very high densities
of energy are present $\left\langle E\right\rangle \sim 300 MeV/fm^3$~\cite{QCD_vac_ChPT}. These values
are only accessible in very dense plasmas (for example in neutron stars~\cite{neutron}) or for very high
fluxes of high energy radiation. For example for radiation energies in the TeV range propagating in vacuum
are required fluxes over $10^{56}\,photons\,m^{-2}s^{-1}$. It is understood that these fluxes include both
the propagating and the background photons.

\section{Vacuum Birefringence}

The relevant radiative corrections that contribute to vacuum birefringence are of
second order in the fine-structure constant~\cite{Birula,Adler}. The usual classical
wave equation in order $\alpha^2$ is linear in the photon field $A$~\cite{Adler}.
Hence for a static transverse magnetic field $B_0$, gathering the results from the previous
section we obtain, due to QCD effects, a correction on the refractive indexes eigenvalues
given by~\cite{Birula,Adler,Gies}
\be
\ba{rcl}
\lambda_\perp&=&\displaystyle 8\,\left(1+\Delta\xi_\mu+\Delta\xi_c+\frac{7}{4}\Delta\xi_{\pi^\pm}\right)\,\xi_e\,B_0^2\ ,\\[5mm]
\lambda_\parallel&=&\displaystyle 14\,\left(\vphantom{\frac{4}{7}}1+\Delta\xi_\mu+\Delta\xi_{\pi^0}+\frac{4}{7}\Delta\xi_{\pi^\pm}\right)\,\xi_e\,B_0^2\ .
\ea
\lb{lambdas}
\ee
The directions $\parallel$ and $\perp$ correspond respectively to the parallel and transverse
directions to the external magnetic field, $\xi_e$ is given in~(\ref{Pi2_ee}) and the
several corrections are ordered by magnitude significance according to the
estimatives given in~(\ref{Pi2_mumu}),~(\ref{Pi2_pi0}),~(\ref{Pi2_pipm}) and~(\ref{Pi2_c}).
The above equations result in having different refractive
indexes for the parallel and perpendicular directions to the magnetic field~\cite{Gies}
\be
N_\parallel=1-\frac{1}{2}\lambda_\parallel\ \ ,\ \ N_\perp=1-\frac{1}{2}\lambda_\perp\ ,
\ee
which introduce a phase shift in the propagating wave. Considering a linearly polarized wave of wave
number $\vb{k}=k_0\,\vb{z}$ which polarization makes an angle $\theta_0$ with a static magnetic field
$B_0$ both gain an ellipticity and its polarization is rotated due to the vacuum effects (see for example~\cite{Gies}).
The polarization rotation is given by
\be
\Delta\theta=\frac{1}{4}\left(\lambda_\parallel-\lambda_\perp\right)\Delta z\,\sin(2\theta_0)\ ,
\lb{rot}
\ee
being $\Delta z$ the distance traveled by the radiation,
and the ellipticity is given by
\be
\psi=-\frac{\omega}{4}\left(\lambda_\parallel-\lambda_\perp\right)\Delta z\,\sin(2\theta_0)\ ,
\lb{ellipse}
\ee
being $\omega$ the radiation frequency. The relative magnitudes of rotation induced in
vacuum by the several effects presented here are pictured in figure~\ref{fig.effects_pol}.
We note that the contribution to the rotation of the pseudo-scalars have the same sign of the
fermionic loops contributions ($\lambda_\parallel>\lambda_\perp$), while the contribution
due to the charged pion loops have the opposite sign ($\lambda^{\pi^\pm}_\parallel<\lambda^{\pi^\pm}_\perp$).
\fig{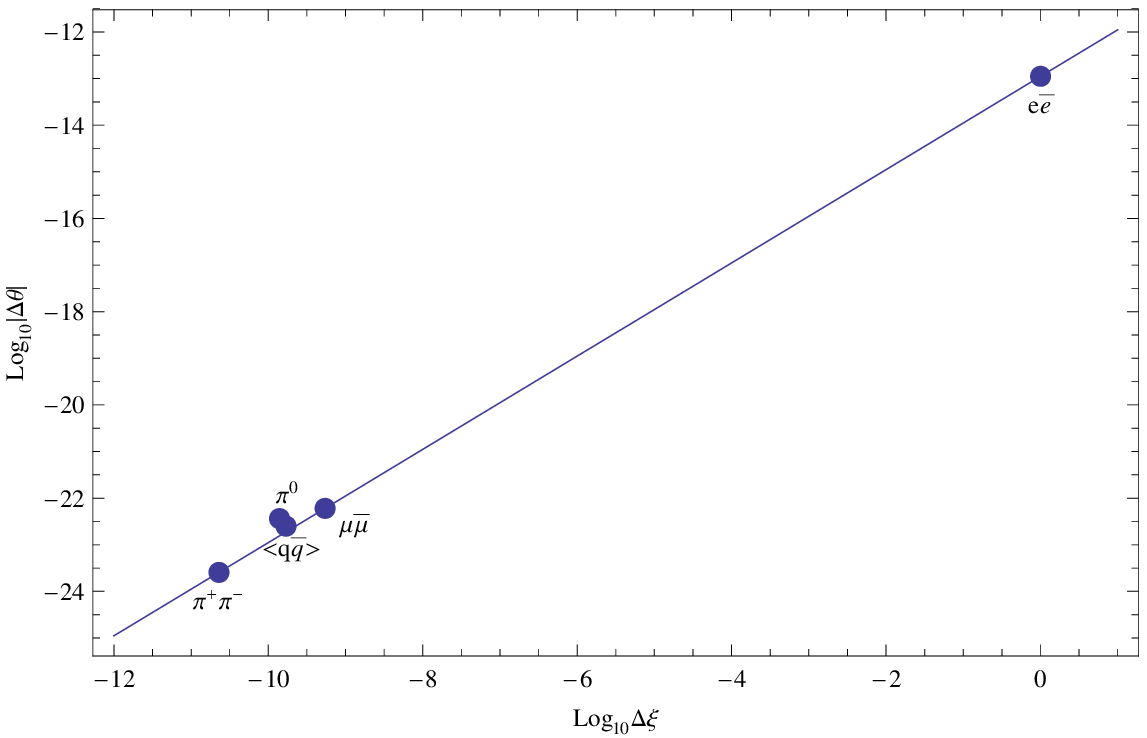}{Relative contributions to the polarization rotation as given by equation~(\ref{rot}) from the several effects as a function of
$\Delta\xi_i$, i.e. each effect magnitude in relation to the magnitude effect due to electron-positron
loops ($e\bar{e}$). Both axes are in logarithmic scale. The continuous line coincides approximately
to the PVLA experimental conditions~\cite{PVLA} with $B_0=5.5\,T$ and $\Delta z=10^9\,m$. The marked
points are labeled and correspond to the QED corrections due to electron-positron loops ($e\bar{e}$),
the muon-antimuon loops ($\mu\bar{\mu}$) interchange of the neutral pion ($\pi^0$), quark condensates ($\left\langle q\bar{q}\right\rangle$)
and charged pion loop ($\pi^+\pi^-$). The ellipticities are obtained by re-scaling these results by the radiation frequency $\omega$.}{fig.effects_pol}

We note that the usual QCD scale is set by $\Lambda_{QCD}\approx 200\,MeV$, however as already
discussed, we also know that for energies of approximately $\Lambda_q=600\,MeV$ the strong running coupling
constant $\alpha_s$ is of order of unity and the perturbative regime of QCD is no longer valid~\cite{QCD}.
Therefore the correct value of the cut-off corresponding to low-energy quark condensate is not exactly
known and should be in the range $200<\Lambda<600\,MeV$. This value should correspond to the chiral phase
transition energy of the Nambu-Jona-Lasinio theory~\cite{NLJ}.

As already put forward in~\cite{axion2} (see also~\cite{axion3}) these effects, in particular the contribution
of the pion, is several orders of magnitude lower than the one due to the electron vacuum oscillation as well as
of the theoretical axion contribution. In addition we recall that the quark condensate contribution is only
present if very high density of radiation is considered.

We conclude that vacuum birefringence due to QCD corrections is negligible for all known
physical systems. As already put forward before in~\cite{axion2}, the main contribution is due
to the neutral pion being many orders of magnitude below the vacuum polarization effects of
virtual electron loops which, although a well established phenomena within QED, has not directly
been detected any polarization rotation, neither in laboratory experiments, neither in
astrophysical environments. However in the next section we give an example where the light scattering
by the $\pi^0$ meson may have measurable effects.

\section{High Energy $\gamma$-ray Propagation}

We can also apply the results derived so far to the propagation
of $\gamma$-ray burst in background magnetic fields where the the
effects studied in the previous sections seem to be relevant due to the
very high radiation energy. The main contribution to the attenuation
of the $\gamma$-ray spectrum for high energies ($E$ above $10^{16}$~eV) is photon
desintegration (particle-antiparticle pair production) due to interaction with
the background electromagnetic fields~\cite{desintegration}. These effects result in a
exponential decaying law for the $\gamma$-ray spectrum, $\sim E^{-\Gamma(z)}$.
For radiation from the center of the galaxy (corresponding to $z=8.5$\,kpc)
the value of the decaying exponent is $\Gamma\approx 2.25$~\cite{HESS}.
The radiation flux is of order $10^{-8}\,photons\,m^{-2}\,s^{-1}\,TeV^{-1}$~\cite{gamma_conv}
(corresponding to a radiation density of order $\rho_\gamma\sim 10^{-54}\,MeV/fm^3$). Therefore
for this particular case, the only relevant contribution discussed in this work is from the
$\pi^0$ meson which we address next.

We note that it is also expected that the axion-like pseudo-scalar contribution has visible effects in
the high energy range (of order of TeV) either increasing or decreasing the optical dept
depending on the values of the mass and photon coupling constant considered. More specifically
it is expected that for more stable particles with long decaying time (low decaying rate) as the
light axion, the optical depth increases~\cite{axion2,pseudo_ray,Raffelt}, while for less stable particles
with lower decaying time (greater decaying rate), the optical depth decreases.
What distinguishes between these cases is the relation of the pseudo-scalar mass ($m_\phi$)
to the photon pseudo-scalar coupling ($g_{\gamma\phi}$) properly taking in consideration the
background and traveling radiation energy. We will return to this discussion
by the end of this section. Specifically the equation for a generic pseudo-scalar $\phi$ mixing
with the photon is~\cite{axion2,Raffelt,pseudo_ray}
\be
\left(\omega-i\partial_z+M\right)\left[\ba{c}A_\parallel\\[3mm]A_\perp\\[3mm]\phi\ea\right]=0
\ee
with
\be
M=\left[\ba{ccc}
\Delta_{\gamma\gamma}+\Delta_\parallel&0&\Delta^\parallel_{\gamma\phi}\\[3mm]
0&\Delta_{\gamma\gamma}+\Delta_\perp&\Delta^\perp_{\gamma\phi}\\[3mm]
\Delta^\parallel_{\gamma\phi}&\Delta^\perp_{\gamma\phi}&\Delta_{\phi}
\ea\right]\ ,
\lb{matrix}
\ee
and the several entries given by
\be
\ba{rclcrcl}
\Delta_{\gamma\gamma}&\approx&\displaystyle -i\frac{\Gamma}{2z_0}\ln\left(E\right)&,&\\[5mm]
\Delta_\parallel&\approx&4\xi_e B^2&,&\Delta_\perp&\approx&7\xi_e B^2\ ,\\[5mm]
\Delta^{\parallel,\perp}_{\gamma\phi}&=&\displaystyle\frac{1}{2}g_{\gamma\phi}B^{\parallel,\perp}&,&\Delta_{\phi}&=&m_{\pi_0}\ .
\ea
\ee
The approximation in $\Delta_{\gamma\gamma}$ corresponds to a linearization of the cross section
for photon desintegration in the TeV range of the energies ($E$) for distances $z_o\approx 8.5$~kpc and
we have considered the photon mass negligible ($m_\gamma\approx 0$).
The approximation in $\delta_{\parallel,\perp}$ corresponds to neglecting the contributions from $\mu\bar{\mu}$ loops
and the charged pion $\pi^+\pi^-$ loops. As derived in~\cite{Raffelt}, for non-polarized radiation in
gaussian magnetic field distributions, the conversion probability of photons to pseudo-scalars is
\be
\ba{rcl}
P_{\gamma\to\phi}&=&\displaystyle\frac{1}{3}\left(1-e^{-\frac{3P_0z}{2s}}\right)\ ,\\[5mm]
P_0&\approx&\displaystyle 0.4\times 10^{-7}\left(\frac{g\,B_G E_{10}}{m_\phi^2}\right)^2\ .
\ea
\lb{prob}
\ee
These expressions are applicable to the mixing of the neutral pseudo-scalars to photons in $\gamma$-ray bursts with
coupling given by $g=g_{\gamma\phi}/10^{-6}\,GeV^{-1}$, the root mean square magnetic field strength $B_G=1\,\mu Gauss$,
the energy $E_{10}$ given in units of $10\,TeV$ and the mass given in $MeV$. $z$ is the distance to the source in
$pc$ and $s$ the size of the magnetic field domains also in $pc$~\cite{Raffelt}. For $\gamma$-ray burst from the center of the
galaxy one has $z=8.5\,kpc$ and $s=0.01\,pc$~\cite{HESS,Raffelt}. For the particular cases of the $\pi_0$ mixing we
have $g=2.49\times 10^4$ and $m=135\,MeV$. The deviation to the power law considering this effect is pictured in
figure~\ref{fig.effects_sep}.
\fig{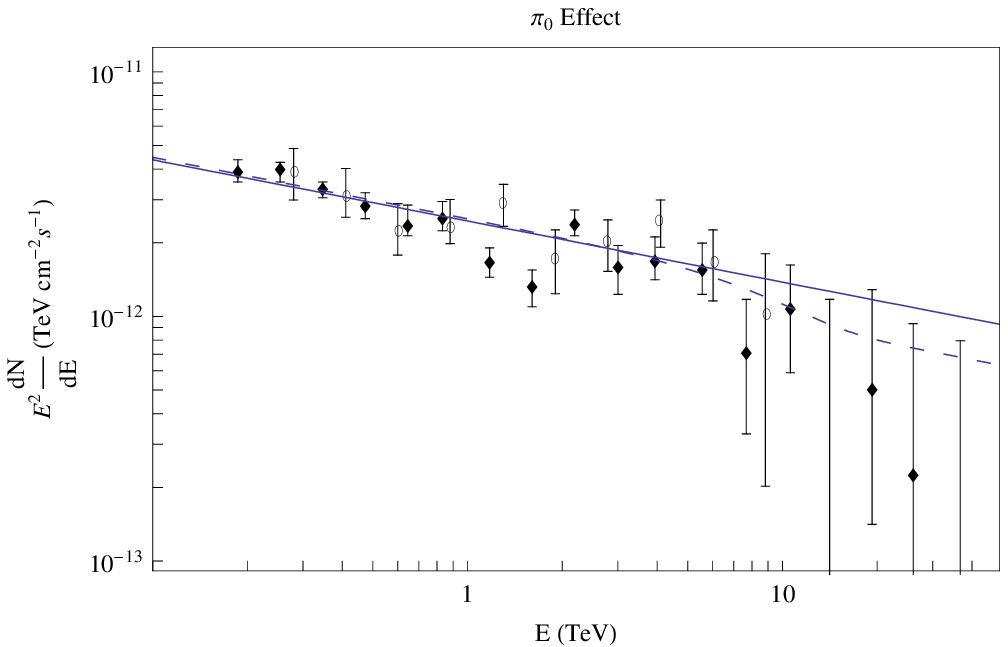}{Deviations from the power law due to $\pi_0$ exchange. The open circles and filled circles are
data points from HESS collaboration corresponding to the data sets from July/august of~2003 and~2004~\cite{HESS}
(see also~\cite{Raffelt}). The filled line represents the power law best fit $dN/dE\sim E^{-\Gamma}$ with
$\Gamma=2.25$ and the dashed line to the contribution of the $\pi_0$.}{fig.effects_sep}

As we have put forward in the beginning of this section, depending on the specific coupling constants, masses and
existing energies the effects of the pseudo-scalar mixing to photons renders quite different results. The relevant
expressions to compare correspond to the diagonal and off-diagonal components of the matrix for pseudo-scalar photon mixing.
Respectively the relevant expressions to compare are $m_\phi^2 z/s$ and $g_\phi B E$. This analysis was originally
carried in~\cite{axion2} and also considered in~\cite{pseudo_ray,Raffelt}. The main differences between this several
works are the numerical values for the parameters of the propagation equation. While
in~\cite{axion2} the length $z/s$ traveled by radiation is of order of Km, in~\cite{Raffelt} are considered astrophysical
environments traveling several kpc which allows to saturate the scalar-photon oscillations as given by
equation~(\ref{prob}). The quantitative similarities of our results in relation to~\cite{Raffelt} is simply due
to the ratios of the couplings to mass squared of the \textit{heavy} axion considered there and the pion considered here,
being different only by about one order of magnitude $21\times g_\pi/m_\pi^2\approx g_{axion}/m_{axion}^2$.
As for~\cite{pseudo_ray} we note that it is considered a very light axion $m_{axion}^2\ll g_{\gamma\,axion} E B$ such that the
massless limit $m_{axion}\to 0$ is taken. This last case is clearly not applicable to the pion where we have that $m_\pi^2> g_{\gamma\pi} B E$
and explains why for a relatively heavy intermediate pseudo-scalar (with low decaying times) the optical-depth is decreased while for a very
light intermediate pseudo-scalar (with high decaying times) the optical-depth in increased.

It could also be considered the more generic case of several pseudo-scalar and scalar mixing. Not only the
theoretically suggested axion, as well as the quark condensate effects when are met the conditions for their
formation (for example close to neutron stars and magnetars~\cite{neutron}).

\section{Conclusions}

In this work we have computed the QCD corrections to QED vacuum polarization effects.
Although we conclude that the contribution to vacuum birefringence of the effects presented
here are negligible when compared to the effect of virtual electron loops they have observable
consequences for high energy $\gamma$-ray propagation. In particular we have shown that the neutral
pion mixing with photons significantly contribute to a deviation from the power-law spectrum in the $TeV$
range that may be relevant when considering the superposition of other pseudo-scalar effects in this
range~\cite{Raffelt,gamma_conv}. As for quark condensates and virtual quark loops we deduced that only
for very high radiation energy fluxes ($\rho_c > 300\,MeV/fm^3$) and strong
magnetic fields ($B > 10^{14}\,T$), their effects may be relevant. Hence near neutron stars and
magnetars~\cite{neutron} these effects may affect $\gamma$-ray polarization~\cite{gamma_pol}.

\ \\
{\large\bf Acknowledgments}\\
The authors thank Jos\'e Tito Mendon\c{c}a who contributed to the development of the previous version
of this work. Would also like to thank Jo\~ao Seixas, Michael Scadron, George Rupp as well as to
Em\'ilio Ribeiro for several discussions, in particular by calling our attention to the fact that
$\gamma$-ray do not have the required energy density to produce quark condensates. We also thank the
referee for valuable comments and suggestions. Work of PCF supported by SFRH/BPD/17683/2004 and SFRH/BPD/34566/2007.

\end{document}